\documentclass[intlimits,twoside,a4paper]{article}

\usepackage{amsmath,amssymb}
\usepackage{graphicx}
\usepackage{wrapfig}

\usepackage[T2A]{fontenc}
\usepackage[cp1251]{inputenc}
\usepackage[eqsecnum]{cmpj}
\issue{2011}{14}{4}{43002}

\doinumber{10.5488/CMP.14.43002}

\title[Free energy of an Ising-like model]%
{Gibbs free energy and Helmholtz free energy\\ for a three-dimensional Ising-like model}

\author[M.P. Kozlovskii, R.V. Romanik]{M.P. Kozlovskii, R.V. Romanik}
\address{Institute for Condensed Matter Physics of the National Academy of Sciences of Ukraine,
\\1 Svientsitskii Str., 79011 Lviv, Ukraine}

\date{Received October 12, 2011, in final form November 15, 2011}

\authorcopyright{M.P. Kozlovskii, R.V. Romanik, 2011}

\begin{document}

\maketitle

\begin{abstract}
The critical behavior of a $3D$ Ising-like system is studied at
the microscopic level of consideration. The free energy of
ordering is calculated analytically as an explicit function of
temperature, an external field and the initial parameters of the
model. Within a unified approach, both Gibbs and Helmholtz free
energies are obtained and the dependencies of them on the external
field and the order parameter, respectively, are presented
graphically. The regions of stability, metastability, and
unstability are established on the order parameter--temperature
plane. The way of implementation of the well-known Maxwell
construction is proposed at microscopic level.
\pacs 05.50.+q, 05.70.Ce, 64.60.F-, 75.10.Hk
\end{abstract}

\section{Introduction}

Basic principles of phenomenological theory for second order phase
transitions (PT) were formulated by Landau in late 30--40s~\cite{Dau1_37,Dau2_37}, originally to describe superconductivity.
The key assumption of the theory is that in the vicinity of the
critical point, the free energy can be expanded  in a power series
in the order parameter, the equilibrium value of which is found to
obey the minimum condition for free energy. As is now known~\cite{Land_05}, this expansion is well defined in the neighborhood
of $T_{\mathrm c}$ (critical value of temperature) except for a narrow
interval determined by the Ginzburg criterion~\cite{Ginz_60}. In
this interval, the crucial role is played by the order parameter
fluctuations while Landau's theory is essentially a mean-field
theory. Nevertheless, it is of great use as a qualitative tool for
understanding the nature of PT.

The above mentioned power series expansion is often called
Landau's free energy. It turns out that similar quantity can also
be derived (not constructed) in a theory considering PT at a
microscopic level. Indeed, in the collective variables (CV) method
proposed in~\cite{Yukh_87,Y_Nuovo} to describe the critical
behavior of spin systems, such a quantity appears in a natural
way. In particular, a microscopic analogue of Landau free energy
was calculated in~\cite{PRB_02}, where the Ising-like system was
considered in zero external field.

The purpose of the present paper is to extend the results obtained
earlier by the CV method for physical characteristics of 3D
Ising-like model to the presence of an external field  and look at
what is going on when the field changes sign. As a result, we
obtain the Gibbs free energy, Helmholtz free energy, and the order
parameter and establish the regions of stability, metastability,
and unstability for the system under investigation.

\section{Method}

In our research we consider the system of $N$ Ising spins placed
on sites of a simple cubic lattice of the spacing  $c$. The
Hamiltonian of such a system in an external field is well known
\begin{equation}
\label{H}
  H=-\frac12\sum_{{i},{j}}\Phi(r_{{i},{j}})\sigma_{{i}}\sigma_{{j}}
  -{\cal{H}}\sum_{{i}}\sigma_{{i}}\,.
\end{equation}
Here, \(\Phi(r_{{i},{j}})\) is a short-range
interaction potential between spins located at the $i$-th and
$j$-th sites, the spin variables $\sigma_{{i}}$ take on
values $\pm 1,$ ${\cal H}$ is the external field. We do not
restrict the summation in (\ref{H}) to the nearest neighbors. To
this end, $\Phi(r)=\text{const}\times\exp{(-r/b)}$ can be chosen
as the interaction potential with an effective range~$b$.

It is also known that this problem has not yet been solved.
Universal critical characteristics of three-dimensional (3D)
systems are successfully described in a variety of approaches.
Among them it is worth mentioning the field theoretical and
renormalization group (RG) methods~\cite{Zinn_J_07}, which also
enable one to calculate nonuniversal characteristics (critical
amplitudes, critical temperatures), but without taking into
account the dependency on initial parameters of the Hamiltonian.
An alternative method for theoretical investigation of the model
under consideration is the method of CV~\cite{Yukh_87}, which has
an advantage of being a successive microscopic approach, which, in
turn, makes it possible for thermodynamic functions and physical
characteristics of the system to be explicitly obtained as
functions of temperature, the field, and microscopic parameters of
the model.

In the framework of the ``$\rho^4$-model'' approximation the
functional representation for the partition function in terms of
CV $\rho_{\textbf{k}}$ is as follows:
\begin{eqnarray}
\label{Z_func}
Z&=&Z_0\int(d\rho)^{N_0}\exp
\bigg[a_1\sqrt{N_0}\rho_0-\frac{1}{2}\sum_{\textbf{k}\in{\cal{B}}_0}d(k)\rho_{\textbf{k}}\rho_{-\textbf{k}}-\nonumber\\
&-&\frac{a_4}{4!}N_0^{-1}\sum_{\textbf{k}_i\in{\cal{B}}_0}
\rho_{\textbf{k}_1}\ldots\rho_{\textbf{k}_4}\delta_{\textbf{k}_1+\ldots+\textbf{k}_4}
\bigg].
\end{eqnarray}
Here, the quantity $d(k)$ contains the Fourier transform of the
interaction potential
\begin{equation}
\label{d_k}
d(k)=a_2+\beta\Phi(0)\bar\Phi-\beta\Phi(k).
\end{equation}
The explicit expressions for \(a_n\) can be found in~\cite{CMP_05}
(equation~(1.25)) , $N_0=N/s_0^3$ where the quantity $s_0$ defines the
region of validity for the parabolic approximation of Fourier
transform of the interaction potential (for details, see
\cite{CMP_10}). The summation in (\ref{Z_func}) is performed over
wave vectors of the first Brillouin zone corresponding to a
reciprocal effective lattice with the lattice constant $cs_0$.

In the work~\cite{CMP_09} the method for calculation of the
partition function (\ref{Z_func}) of the model near the second
order PT point was generalized to the case of the presence of the
fixed external field of an arbitrary magnitude. Unlike the work
\cite{PRB_06}, here no assumption concerning either the strength
or the weakness of the applied field is made and, therefore,no
perturbation series are used. The calculation procedure is based
on Kadanoff's idea of constructing effective spin blocks
\cite{Kad_66}. We divide the phase space of the CV
$\rho_{\textbf{k}}$ into layers, the RG parameter being $s$, and
average the Fourier transform of the potential in each $n$-th
layer. The first step in calculating equation~(\ref{Z_func}) is to
integrate over $\rho_{\textbf{k}}$ with $k\in
[k_{\mathrm {max}},k_{\mathrm {max}}/s],$ $k_{\mathrm {max}}={\pi}/{cs_0}.$ The result of
integration proves to be represented in the form of the initial
expression, equation~(\ref{Z_func}), with renormalized coefficients
$a_1^{(1)},$ $d_1(k),$ and $a_4^{(1)}.$ If we perform a
step-by-step integration of the partition function over $n_p$
layers, we arrive at
\begin{equation}
\label{Z3}
Z=Z_0\big[Q(d)\big]^{N_0}\bigg(\prod_{n=1}^{n_p}Q_n\bigg)Z_{\rm LGR}\, .
\end{equation}
The partial partition functions $Q_n$ of the $n$-th level are
characterized by a set of coefficients $a_1^{(n)},$ $d_n(0)$,
$a_4^{(n)}$, for which the recurrence relations (RR) hold (the
work~\cite{CMP_05} is devoted to this problem). The quantity $n_p$
is called the exit point from the critical regime of the order
parameter fluctuations. It defines the number of iterations at
which the system is still in the scaling region.

In~\cite{PRB_06}, two regions in $(h,\tau)$-plane were
distinguished, which are of weak and of strong field. The
calculations in the regions were performed using different forms
of exit points. These quantities were subject to certain
conditions which we do not discuss in the present work. If the
field was considered to be strong, the formula
$n_p=n_h=-\ln{\tilde h}/\ln{E_1}-1$ was used, while in the case of
a weak field another expression
$n_p=m_{\tau}=-\ln{\tilde{\tau}}/\ln{E_2}-1$ was taken. Hence, the
question immediately arises what should be done at intermediate
values of a field. The problem is solved by constructing a new
expression for the exit point, proposed in~\cite{CMP_09}:
\begin{equation}
\label{EP_old}
n_p=-\frac{\ln{\left(\tilde{h}^2+{h_{\mathrm c}^{(\pm)}}^2\right)}}{2\ln{E_1}}-1,
\end{equation}
where some temperature fields $h_{\mathrm c}^{(+)}=|\tau|^{p_0}$ and
$h_{\mathrm c}^{(-)}=|\tau_1|^{p_0}$ are introduced, $p_0=\ln{E_1}/\ln{E_2}$
is the so-called ``gap exponent''~\cite{Stanley_71}, the signs ``+''
and ``--'' are related to $T>T_{\mathrm c}$ and $T<T_{\mathrm c}$ respectively, and
\begin{equation}
\tilde{h}=\frac{s_0^{3/2}}{h_0}h, \quad h=\beta{\cal{H}},
\quad\quad \quad\tilde{\tau}=\frac{c_{k_1}}{f_0}\tau , \quad
\tau_1=-E_2^{n_0}\tau, \quad \tau=\frac{T-T_{\mathrm c}}{T_{\mathrm c}}\,.
\end{equation}
The quantities $E_1$ and $E_2$ are the eigenvalues of the matrix
of the RG transformation linearized near the fixed point of RR.

In the limiting cases, as $h\rightarrow 0$ or as $\tau\rightarrow
0,$ equation~(\ref{EP_old}) takes on the form of $m_\tau$ or $n_h,$
respectively, and therefore can be applied to the case of an
arbitrary field. Nevertheless, it is worth mentioning that the
inequality $\tilde{h}\gg h_{\mathrm c}$ defines a strong field, and
$\tilde{h}\ll h_{\mathrm c}$ defines a weak field. Note also that there is
an arbitrariness in choosing $n_p,$ which is discussed in
\cite{CMP_09,CMP_10} in more detail. What is important is that for
$n>n_p$ one should be able to integrate the partition function
with Gaussian measure density.

\begin{table}[h]
\caption{Numerical values of the parameters used in present
calculations.} \label{tab_coefs}
\begin{center}
\begin{tabular}{|c|c|c|c|c|c|c|c|c|c|}
%\begin{tabular}{cccccccccc}
\hline $b/c$ & $\beta_{\mathrm c}\Phi(0)$ & $s_0$ & $f_0$ & $h_0$ & $c_{k_1}$ & $\phi_0$ & $\Phi_f$ & $n_0$ \\
\hline\hline
0.3 & 1.6411 &2.0  &0.5& 0.760 & 1.1762 &0.5938 & 0.105 & 0.5\\
\hline
\end{tabular}
\end{center}
\end{table}

We will adhere to the approach adopted in~\cite{UFZ_09}, where the
coefficients $c_{k_1},$ $f_0,$ $h_0$ are also defined (see equations~(4.4), (4.17), (4.23), and (4.49) in~\cite{UFZ_09}), and take
$E_1=24.551,$ $E_2=8.308.$ The information on the physical meaning
of coefficient $n_0$ can be found in~\cite{JPS_09}. Numerical
values of all coefficients used to obtain the final graphical
results are presented in table~\ref{tab_coefs}. It is worth
stressing that having fixed the value of RG parameter $s=s^*$
($s^*=3.5977$ in the present calculation), there remains only one
initial parameter, which is the ratio of the effective range of
interaction $b$ to the lattice constant $c$. All the other
coefficients can be expressed via $b/c$~\cite{CMP_10,Mic_Theory_01}.

\section{Results and discussion}

Based on the above mentioned method, in the work~\cite{CMP_09} as
well as in~\cite{JPS_09} free energy of 3D Ising-like model was
calculated as a logarithm of the partition function (\ref{Z3})
multiplied by $-kT$ and presented in the form of several
contributions
\begin{equation}
\label{free_en}
F(\tau,h)=F_a+F_s^{(\pm)}+F_0^{(\pm)}.
\end{equation}
Here, the term $F_a$ is the analytical part of free energy and
does not affect the critical behavior of the system. The last two
terms in r.h.s. of (\ref{free_en}) have non-analytical dependence
on temperature and on the external field. The explicit expressions
for $F_s^{(\pm)}$ and $F_0^{(\pm)}$ are
\begin{equation}
\label{Fs}
F_s^{(\pm)}(\tau,h)=-kTN\gamma_s^{(\pm)}\left(\tilde{h}^2+{h^{(\pm)}_{\mathrm c}}^2\right)^{\frac{d}{d+2}},
\end{equation}
and
\begin{equation}
\label{F0}
F_0^{(\pm)}(\tau,h)=-kTNE_0(\sigma_{\pm}).
\end{equation}
The quantity $\gamma_s^{(\pm)}$ from (\ref{Fs}) includes
contributions from the critical regime of the order parameter
fluctuations and from the limiting Gaussian regime (LGR)(for
details, see~\cite{CMP_09,JPS_09}). The mentioned critical regime
is characterized by the RG symmetry.

The quantity $E_0(\sigma_{\pm})$ is the contribution from the
collective variable $\rho_0$. As is known from the theory of CV
\cite{Yukh_87}, the mean value of $\rho_{\textbf{k}=0}$ is
connected with the order parameter and consequently the quantity
$F_0$ is the free energy of ordering~\cite{PRB_02}. If there is no
external field, $F_0$ is analogous to Landau's free energy in
phenomenological theory of second order PT~\cite{PRB_02}. Since we
work in the ensemble of $N$ particles, with field $h$ and
temperature $\tau$ being independent variables, this analogy is
now not direct. Following Stanley~\cite{Stanley_71}, and based on
convincing arguments of the work~\cite{JMMM_03}, we associate the
partition function of our model with Gibbs free energy. An
appropriate thermodynamic potential is Gibbs free energy provided
the magnetic field and temperature are independent (``natural'')
variables. If the role of independent variables is played by an
order parameter (here magnetization per spin) and temperature,
then the thermodynamic potential associated with the partition
function is Helmholtz free energy. Hence, Landau's free energy is
rather Helmholtz than Gibbs one. The expressions (\ref{free_en})--(\ref{F0}) in turn present Gibbs free energy for the considered
model, although this was not specified in earlier works,
particularly in~\cite{CMP_09,JPS_09}. Therefore, in order to
obtain a microscopic analogue of Landau's energy one should
perform the Legendre transformation. Let $A_0$ denote this
analogue. Then,
\begin{equation}
\label{analogue}
A_0(\tau,M)=F_0+M_0{\cal{H}},
\end{equation}
where $M_0$ is the magnetization of the system under consideration
such that
\begin{equation}
\label{M0}
M_0=-\frac{1}{N}\frac{\partial{F_0}}{\partial{\cal{H}}}=\frac{\partial{E_0}}{\partial{h}}\,.
\end{equation}

The expression for $E_0(\sigma_\pm)$, which was found in
\cite{CMP_09} (see equation~(4.4) there) for $T>T_{\mathrm c}$ and in
\cite{JPS_09} (see equation~(4.3) there) for $T<T_{\mathrm c},$ reads
\begin{equation}
\label{E01}
E_0(\sigma_{\pm})=h\sigma_{\pm}-\frac{1}{2}d_{n_p+2}(0)
\sigma_{\pm}^2-\frac{s_0^3s^{3(n_p+2)}}{4!}a_4^{(n_p+2)}\sigma_{\pm}^4\,.
\end{equation}
The quantity $\sigma_{\pm}$ was found from condition
\begin{equation}
\frac{\partial{E_0(\sigma_{\pm})}}{\partial{\sigma_{\pm}}}=0
\end{equation}
in the form
\begin{equation}
\label{sigma}
\sigma_{\pm}=\sigma^{(\pm)}_0s^{-(n_p+2)/2}.
\end{equation}
This results in the cubic equation for $\sigma^{(\pm)}_0$
\begin{equation}
\label{cub_eq}
\sigma_0^3+p\sigma_0+q=0
\end{equation}
with coefficients
\begin{eqnarray}
\label{q_p}
p&=&6s_0^{-3}r_{n_p+2}/u_{n_p+2}\,,\nonumber\\
q&=&-6s_0^{-9/2}s^{5/2}\frac{h_0}{u_{n_p+2}}
\frac{\tilde{h}}{\left(\tilde{h}^2+{h^{(\pm)}_{\mathrm c}}^2\right)^{1/2}},
\end{eqnarray}
where the denotations $r_n$ and $u_n$ are connected with $d_n(0)$
and $a_4^{(n)}$ through
%\begin{equation}
$d_n(0)=s^{-2}r_n$,  $a_4^{(n)}=s^{-4}u_n,$
%\end{equation}
and according to~\cite{CMP_09,JPS_09} are expressed as
\begin{eqnarray}
\label{rnp2}
r_{n_p+2}&=&\beta_{\mathrm c}\Phi(0)f_0\left(-1\pm \tilde{\tau}E_2^{n_p+2}\right),\nonumber\\
u_{n_p+2}&=&[\beta_{\mathrm c}\Phi(0)]^2\phi_0\left(1\pm \tilde{\tau} E_2^{n_p+2}\Phi_f\right).
\end{eqnarray}
Note, that the coefficients $p$, $q$ should be marked by the
superscript $\pm$ (as should be done for $n_p$ from
(\ref{EP_old})), but where it does not cause confusion, we drop
this superscript out. One should just remember about a difference
in temperature scales between the cases of $T>T_{\mathrm c}$ and of $T<T_{\mathrm c}.$

\begin{figure}
\centerline{\includegraphics[width=0.8\textwidth,angle=0]{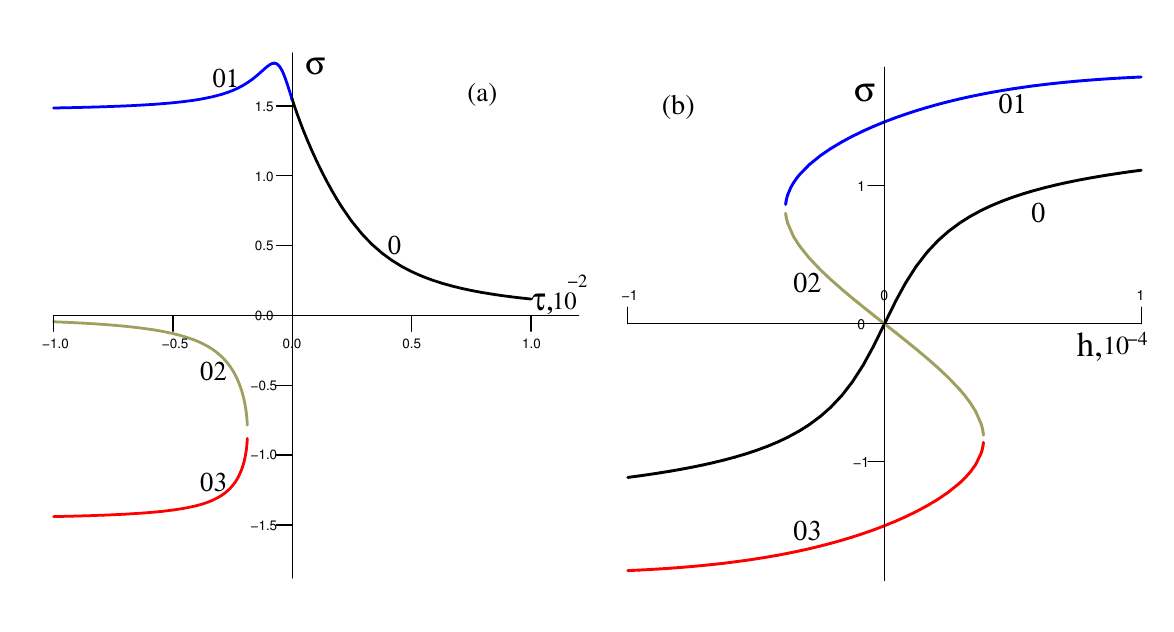}}
\caption{\label{fig_sig}(Color online) Solutions of cubic equation
(\ref{cub_eq}) (a) as functions of temperature for $h=10^{-4}$;
(b) as functions of field for $\tau=\pm 10^{-3}.$ The curves
marked by ``0'' correspond to the real solutions at $\tau\geqslant0;$
marked by ``01'', ``02'', and ``03'' denote different solution at
$\tau<0.$ In this respect, we will refer to the solutions as
$\sigma_0^{(+)}$ and $\sigma_{0i}^{(-)}$ ($i=1,2,3$).
 }
\end{figure}

Equation (\ref{cub_eq}) can be solved by Cardano's method. The
solutions are presented graphically in figure~\ref{fig_sig}. From
the plot we see that for a given value of field there exists some
temperature $\tau_0<0$ such that equation~(\ref{cub_eq}) has three real
solutions for $\tau\leqslant\tau_0$ and one real solution for
$\tau>\tau_0.$ This quantity, $\tau_0,$ corresponds to the
condition $Q=0$ where $Q$ is the discriminant of the cubic
equation~(\ref{cub_eq}):
\begin{equation}
\label{discrim}
Q=(p/3)^3+(q/2)^2
\end{equation}
with $q$ and $p$ from (\ref{q_p}). When $h=0$, then $\tau_0=0.$ If
$h\neq 0,$ then $\tau_0$ is determined by setting r.h.s. of
(\ref{discrim}) equal to zero. In~\cite{JPS_09}, $\tau_0$ was
found numerically and the plot of $\tau_0$ versus $h$ was drawn.
As we will see in what follows, the solution $\tau_0=\tau_0(h)$ to
the equation $Q=0$ defines a curve in ``order
parameter--temperature'' plane which can be identified with the
spinodal of a fluid.

Taking into account (\ref{EP_old}), and (\ref{sigma}), the
quantity $E_0(\sigma)$ takes on
\begin{equation}
\label{E02}
E_0(\sigma)=he_0^{(\pm)}\left(\tilde{h}^2
+{h_{\mathrm c}^{(\pm)}}^2\right)^{\frac{1}{2(d+2)}}-e_2^{(\pm)}\left(\tilde{h}^2+
{h^{(\pm)}_{\mathrm c}}^2\right)^{\frac{d}{d+2}},
\end{equation}
where we have introduced the following notation
\begin{eqnarray}
\label{e0_e2}
e_0^{(\pm)}&=&\sigma_0^{(\pm)}s^{-1/2},\nonumber\\
e_2^{(\pm)}&=&\frac{1}{2}{\sigma_0^{(\pm)}}^2s^{-3}
\left(r_{n_p+2}+\frac{1}{12}u_{n_p+2}s_0^3{\sigma_0^{(\pm)}}^2\right).
\end{eqnarray}

In the previous works~\cite{CMP_09,JPS_09,CMP_10,UFZ_09} the
system was considered in the external field $h\geqslant0$. Analytical
results for (Gibbs) free energy~\cite{CMP_09,JPS_09}, the order
parameter~\cite{JPS_09,CMP_10}, and the susceptibility~\cite{CMP_10} were obtained. The purpose of this paper is to
extend the method to the region $h<0$ and to take into account all
solutions to the equation~(\ref{cub_eq}). We look at Gibbs free
energy $F_0$ and at the contribution from it to the order
parameter $M_0$ of equation~(\ref{M0}). Thereupon, Helmholtz free
energy is calculated.

\begin{figure}%[!t]
\vspace{-0.3cm}
\centerline{\includegraphics[width=0.8\textwidth,angle=0]{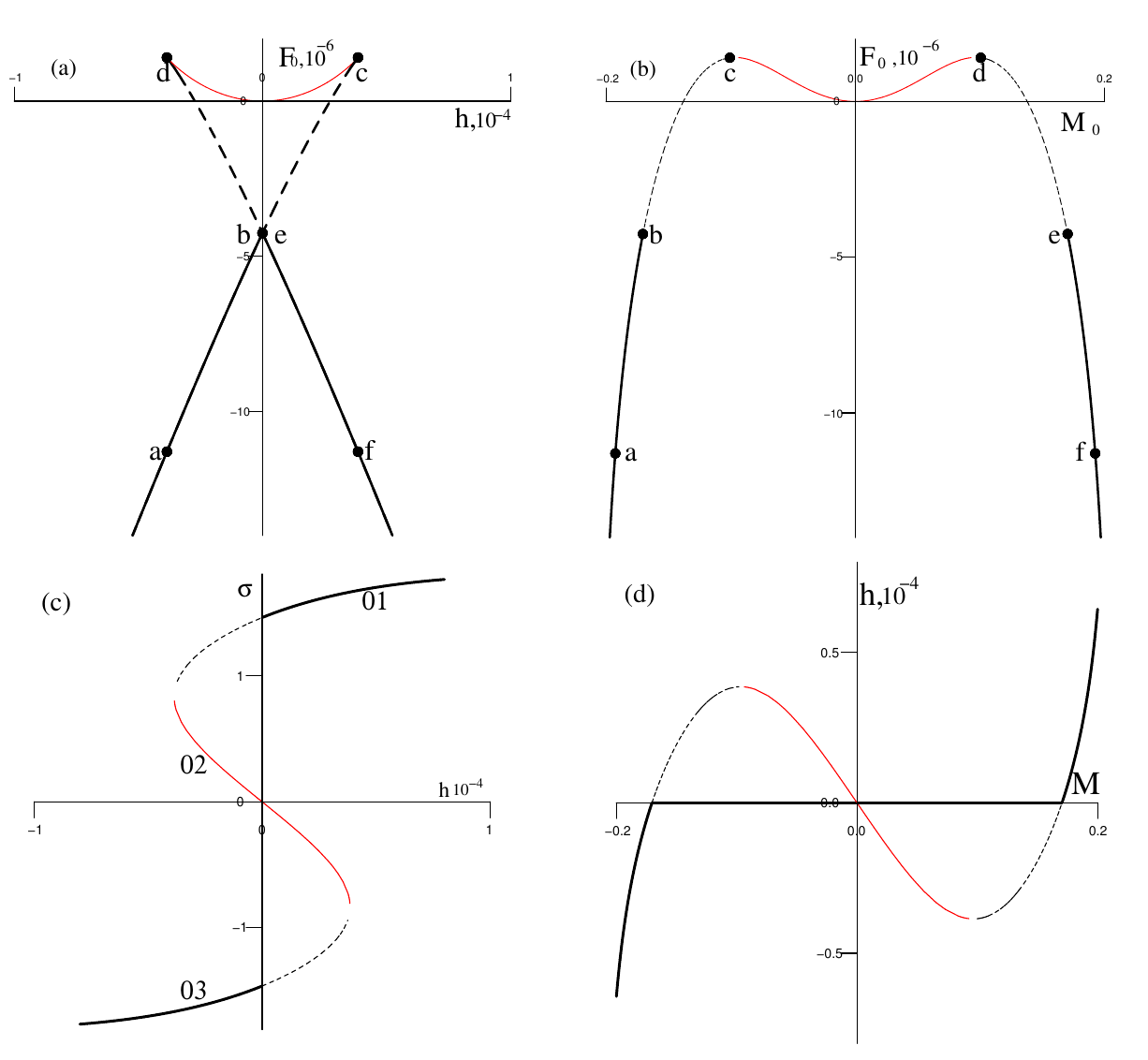}}
\caption{\label{fig_F}(Color online) A set of 4 pictures is placed
for the sake of clarity. All are drawn for $\tau=-0.001.$ The top
left is the field dependence of Gibbs free energy $F_0$. The top
right shows $F_0$ against $M_0.$ The bottom left presents the
solutions of the cubic equation (\ref{cub_eq}). The bottom right
is the equation of state for the system under consideration. The
values of energies are normalized by dividing by $kTN.$ The points
$a$ to $f$ represent some particular states of the system in
different coordinates. See the text for details.}
\end{figure}

\begin{figure}%[!h]
\vspace{0.3cm}
\centerline{\includegraphics[width=0.5\textwidth,angle=0]{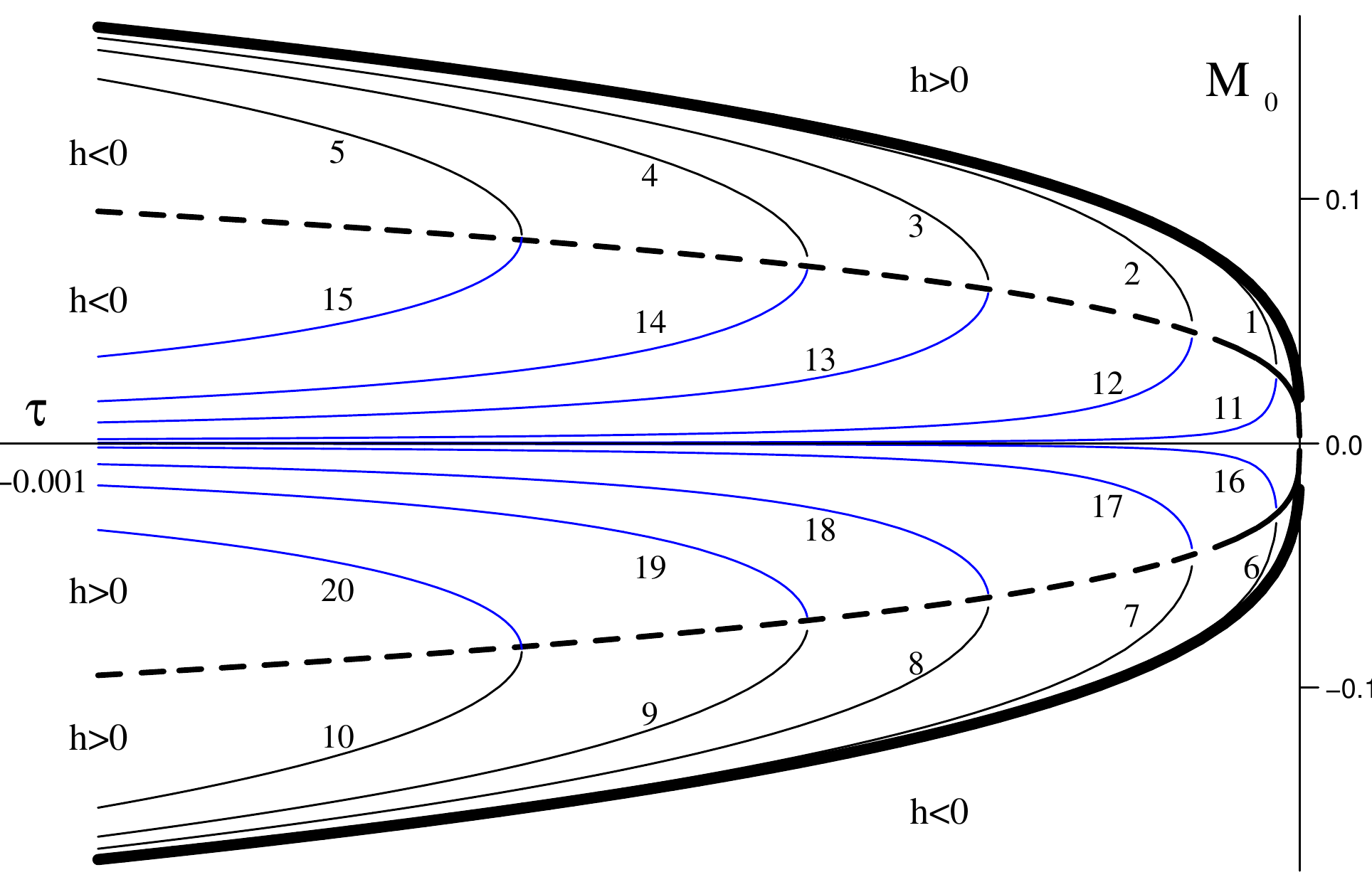}}
\caption{\label{coex}(Color online) The plot shows temperature
dependence of magnetization calculated at different solutions to
equation~(\ref{cub_eq}). Thick line is magnetization in zero external
field and corresponds to the coexistence curve (binodal). In
negative external field, the solution $\sigma_{01}$ gives rise to
Curves 1--5 on the magnetization-temperature plane. In positive
external field, $\sigma_{03}$ gives rise to Curves 6--10. Curves
11--20, which correspond to $\sigma_{02},$ can be interpreted as
non-physical. Thick dashed curve corresponds to the saturation
curve (spinodal). In order to recover that, one should compute
magnetization corresponding to $\sigma_{01}$ at $h<0$ and
$\tau_0=\tau_0(|h|)$ (upper branch) and corresponding to
$\sigma_{03}$ at $h>0$ and $\tau_0=\tau_0(h)$ (lower branch). The
magnitudes of field, at which the results are drawn, are $|h|=0,
10^{-7}, 10^{-6}, 5\cdot 10^{-6}, 10^{-5}, 2\cdot 10^{-5}.$}
\end{figure}

\begin{figure}
\centerline{\includegraphics[width=0.5\textwidth,angle=0]{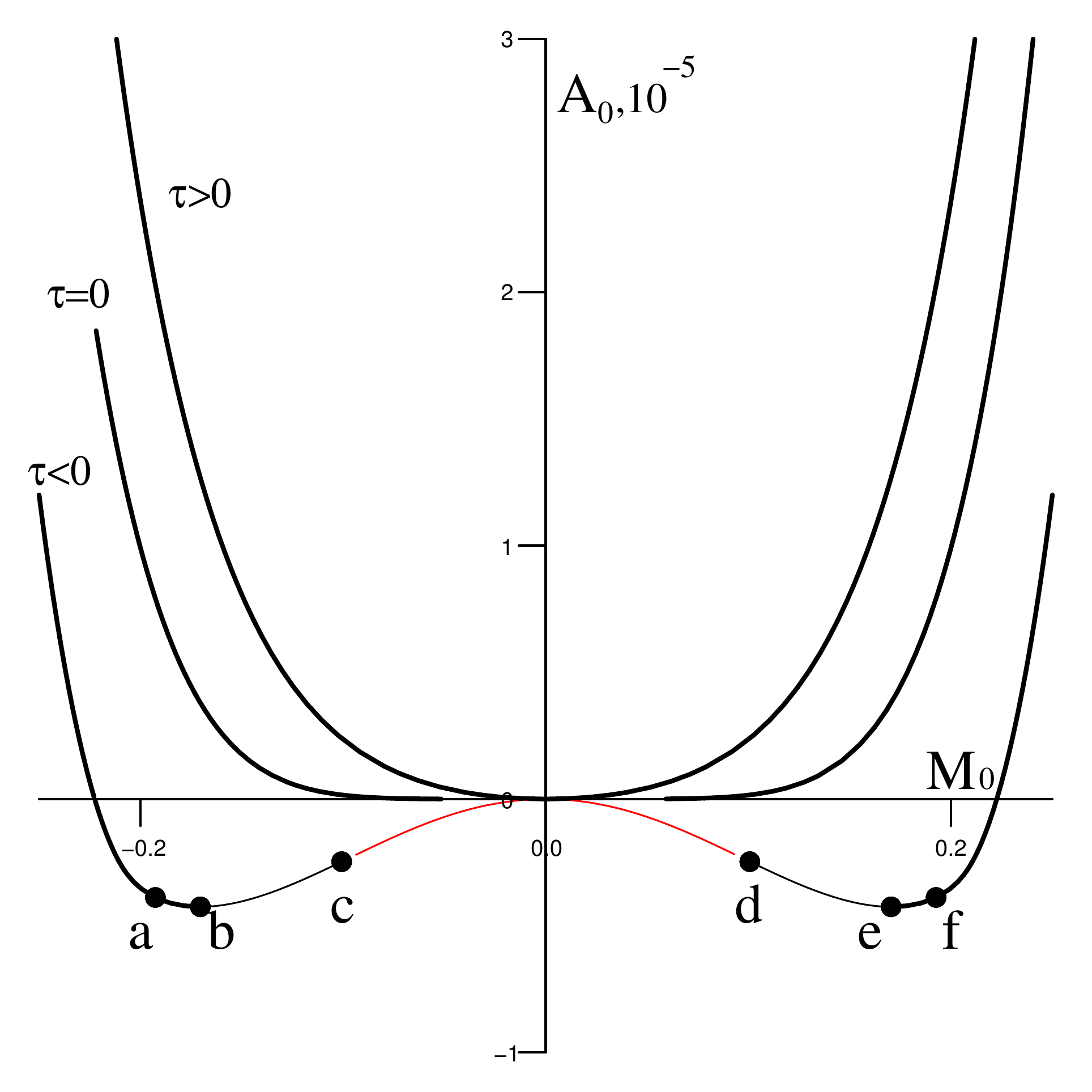}}
\caption{\label{fig_A}(Color online) Helmholtz free energy $A_0$
as a function of order parameter for three different temperatures
$T<T_{\mathrm c}$ ($\tau=-0.001$), $T=T_{\mathrm c}$ ($\tau=0$), and  $T>T_{\mathrm c}$
($\tau=0.001$). The value is normalized by dividing by $kTN$. }
\end{figure}

The main results of this paper are presented in figures~\ref{fig_F},
\ref{coex},~and~\ref{fig_A}. In figure~\ref{fig_F}
 we can see: (a) Gibbs free energy $F_0$
as a function of the external field in low temperature region
($\tau=-0.001$); (b) a plot of $F_0$ versus $M_0$. Each of the
points $a$, $b$, $c,\ldots$ represents a certain specific state of
the system in different coordinates. Especially, points $a$, $c$,
$d,$ and $f$ correspond to such a relation between the field $h$
and temperature $\tau$ that the equality $\tau=\tau_0(h)$ holds.
Thus, in each of these we have $Q=0.$ Points $b$ and $e$ are
associated with first order phase transition and correspond to
$h=0$ and $\tau=-0.001.$ The isotherm in figure~\ref{fig_F}~(b) was
drawn with the help of parametric representation \{$F_0=F_0(h)$,
$M_0=M_0(h)$\}.

Recall that Gibbs free energy is a concave function of field~\cite{Stanley_71}. Therefore, from figure~\ref{fig_F}~(a) we can
conclude
 that the solution $\sigma_{02}^{(-)}$
gives rise to a non-physical result. It corresponds to the region
of unstability ($c-d$ part of the curve in figure~\ref{fig_F}~(a),
(b)). This is not surprising if we note that $\sigma_{02}^{(-)}$
maximizes $F_0.$ Solutions $\sigma_{01}^{(-)}$ at $h<0$ and
$\sigma_{03}^{(-)}$ at $h>0$  correspond to the region of
metastable states ($d-e$ and $b-c$ parts of the curve
respectively). In the case of liquid-gas PT it would be the region
between the binodal and the spinodal curves. Hence, we regard
these results important from the perspective of using this method
for description of the critical behavior in simple fluids.

Figure~\ref{fig_F}~(c) to some extent repeats figure~\ref{fig_sig}~(b)
and is placed here for convenience of representation. The quantity
$\sigma_0$ is related to the scaling function of magnetization via
formula~(4.8) in work~\cite{JPS_09}. Finally, in figure~\ref{fig_F}~(d) the equation of state computed by means of~(\ref{M0}) is
presented at $\tau=-0.001.$

A more detailed representation of the equation of state is shown
in figure~\ref{coex}. The derivatives of Gibbs free energy with
respect to field are presented as functions of temperature,
different solutions to equation~(\ref{cub_eq}) being taken into account.
As we can see, such an accounting allows us to establish spinodal
and binodal curves of the model and, in this respect, to find the
stable, metastable, and unstable regions on ``order parameter--temperature'' plane.

The order parameter dependence of free energy $A_0$ is presented
in figure~\ref{fig_A} for three different temperatures $T>T_{\mathrm c}$,
$T=T_{\mathrm c}$ and $T<T_{\mathrm c}$. Here again $a$, $b$, $c,\ldots$ denote the
states as in figure~\ref{fig_F}. We can see that Helmholtz free
energy has two minima below $T_{\mathrm c}.$ The equilibrium values for the
order parameter are defined by these minima. Knowing their
positions at different temperatures, we can recover the
coexistence curve. Points $b$ and $e$ coincide with the minima at
$\tau=-0.001.$  Note that based on the principle of minimum Gibbs
free energy and from figure~\ref{fig_F}~(a), we can conclude that
with the field changing from $h<0$ to $h>0$ the system will tend
to move along $a-b-e-f$ part of the curve. Applied to Helmholtz
free energy, it means that the system will tend to jump from state
$b$ to state $e.$ Therefore, one may supplement figure~\ref{fig_A}
with a double-tangent construction. As a consequence, on $M_0-h$
plane we have a horizontal segment for $M_0$ at $h=0$ (see
figure~\ref{fig_F}~(d)), which corresponds to a Maxwell construction
derived at microscopic level. On the other hand, based on the
maxima of Gibbs free energy with respect to $M_0$ (see
figure~\ref{fig_F}~(b)), we are able to construct a saturation
curve. Let us just remember that dependence $F_0$ on the order
parameter is formal due to $M_0$ being not ``natural'' variable for
$F_0.$ However, the same results for the binodal and the spinodal
can immediately be obtained if one appropriately chooses  the
solutions to~(\ref{cub_eq}).

The forms of isotherms in figure~\ref{fig_A} are similar to those in
Landau's theory~\cite{Land_05}. However, the peculiar feature is
that in the present work the free energy is an explicit function
of temperature, external field and microscopic parameters (in the
present case a ratio $b/c$) of the model, with a non-analytical
dependence on its arguments. This gives rise to a non-classical
critical behavior of the system, i.e., to the critical exponents
taking on non-classical values. These values for the most
important exponents are reported in table~\ref{tab_exp}. There are
also collected the values obtainable by CV method in higher,
$\rho^6$ approximation as reported in Conclusions of the work~\cite{PRB_02}. Note that in our calculation we have neglected the
critical exponent $\eta$ responsible for the behavior of the pair
correlation function at $T=T_{\mathrm c}.$ Accounting for corrections to
scaling is beyond the scope of the present paper as well. In
table~\ref{tab_exp} the classic values for the critical exponents
are presented as well as the results of other authors. We stick to
the following notation. The temperature behavior of magnetization
is governed by $\beta,$ of the heat capacity by $\alpha,$ of
susceptibility by $\gamma,$ of the correlation length by $\nu.$
The field behavior of magnetization is governed by $\delta$,
$M\sim |h|^{1/\delta}.$ We have calculated some of the exponents
with the help of scaling laws. Based on earlier results~\cite{UFZ_09}, we are also able to estimate the critical exponents
$\varphi$ and $\psi$ that describe the field behavior of the heat
capacity, $C\sim |h|^{-\varphi},$ and of the entropy, $S\sim
|h|^{\psi},$ respectively. Their numerical values are
$\varphi=0.122$ and $\psi=0.539,$ which differ from $\varphi=0$
and $\psi=2/3$ in mean-field theory~\cite{Stanley_71}.

\begin{table}[h]
\caption{Numerical values of the critical exponents. CV,
collective variables method; MF, mean-field values; HT,
high-temperature expansion results; FT, field theory approach.}
\label{tab_exp}
\vspace{2ex}
\begin{center}
\begin{tabular}{|c| c| c| c| c | c|}
%\begin{tabular}{cccccccccc}
\hline
 & CV,  & CV,  & MF & HT~\cite{but_per_11,but_com_05} &FT~\cite{Zin_01} \\
 &{$\rho^4$-model}&$\rho^6$-model~\cite{PRB_02}&\phantom{$\rho^4$-model}&&\\
\hline\hline
$\alpha$ & 0.185 & 0.088  &0& 0.110(1)\phantom{0} & 0.109(4)\phantom{00}   \\
%\hline
$\beta$ & 0.302 & 0.319 & 1/2 & 0.3263(4)  & 0.3258(14)  \\
$\gamma$ & 1.210 & 1.275 & 1 & 1.2373(2)  & 1.2396(13) \\
$\nu$ & 0.605 & 0.637 & 1/2 & 0.6301(2)  & 0.6304(13) \\
$\delta$ & 5 & 5 & 3 & 4.792 & 4.805 \\
\hline
\end{tabular}
\end{center}
\end{table}

It should also be mentioned that the scaling properties for
physical characteristics of the considered model are discussed in
earlier works. In section 3 of the work~\cite{CMP_10} the scaling
functions for (Gibbs) free energy, for the order parameter, and
for susceptibility were calculated.

\section{Conclusions}
In this work, an analytical expression for Gibbs free energy of
Ising-like system is obtained and investigated near the second
order phase transition. The emphasis is made on the presence of
the external field  and on the situation where the field changes
its direction. Some well-defined concepts of phenomenological
theory of phase transitions - e.g. Landau's energy, the Maxwell
construction, the double-tangent construction - are derived and
reexamined at microscopic level. We hope that the approach
described and the results presented should be useful in further
investigations of critical behavior in 3D Ising-like systems with
analytical methods; in particular, in order to obtain the
coexistence curve and the spinodal curve for a system at least in
the vicinity of critical point. There exists a general belief
\cite{Stanley_71,but_per_11,Fish_Zinn_99} that simple fluids
belong to the Ising universality class. Hence, we regard these
results important from the perspective of using this method for
description of the critical behavior in simple fluids.

\newpage

\ukrainianpart

\title{Вільна енергія Гіббса та вільна енергія Гельмгольца для тривимірної ізінгоподібної моделі}

  \author{М.П. Козловський, Р.В. Романік}
  \address{Інститут фізики конденсованих систем НАН України, вул. І.~Свєнціцького, 1, 79011 Львів, Україна }

\makeukrtitle

\begin{abstract}
\tolerance=3000%
В даній роботі на мікроскопічному рівні розгляду вивчається критична поведінка тривимірної ізінгоподібної системи.
Аналітично обчислюється вільна енергія впорядкування як функція температури, зовнішнього поля і початкових параметрів моделі. В межах запропонованого підходу отримано вирази для вільної енергії Гіббса і Гельмгольца, їх залежності від поля і параметра порядку приведені також графічно. На площині параметр порядку--температура знайдені області стабільності, метастабільності та нестабільності. Запропоновано спосіб реалізації правила Максвелла на мікроскопічному рівні.
\keywords модель Ізінга, критична поведінка, зовнішнє поле

\end{abstract}

\end{document}